\documentclass[aps]{revtex4}
\usepackage{amssymb}
\usepackage{amsmath}
\usepackage{bm}
\usepackage{graphicx}
\usepackage{epsfig,amsmath}

\begin{document}

\title{Vortex bundle collapse and Kolmogorov spectrum. Talk given at the Low Temperature Conference, Kazan, 2015}
\author{Sergey K. Nemirovskii\thanks{%
email address: nemir@itp.nsc.ru}}
\affiliation{Institute of Thermophysics, Lavrentyev ave, 1, 630090, Novosibirsk, Russia
and Novosibirsk State University, Novosibirsk}
\date{\today }

\begin{abstract}
The statement of problem is motivated by the idea of modeling the classical
turbulence with a set of chaotic quantized vortex filaments in superfluids.
Among various arguments supporting the idea of quasi-classic behavior of
quantum turbulence, the strongest, probably, is the $k$ dependence of the
spectra of energy,  $E(k)\propto k^{-5/3}$ obtained in numerical
simulations and experiments. At the same time the mechanism of classical vs quantum turbulence (QT) is not clarified and the source of the $k^{-5/3}$ dependence is unclear.
In this work we concentrated on the nonuniform vortex vortex bundles. This choice is related to actively discussed question concerning a role of collapses in the vortex dynamics in
formation of turbulent spectra.  We demonstrate that the nonuniform vortex vortex bundles, which appear in result of nonlinear vortex dynamics generates the energy spectrum, which close to the Kolmogorov dependence $\propto k^{-5/3}$.
\end{abstract}

\maketitle

\section {Introduction and scientific background}

    The problem of modeling classical turbulence with a set of chaotic
quantized vortices is undoubtedly in the mainstream of modern studies of
vortex states in quantum fluids (see, e.g., \cite{Vinen2000},\cite%
{Skrbek2012},\cite{Nemirovskii2013}).\\
    One of the evidences of the quasi-classical behavior of QT is the $k$%
-dependence of the spectra of energy$E(k)$, obtained in numerical simulations
and experiments, and their comparison with the Kolmogorov law $%
E(k)\propto \,k^{-5/3}$. The experimental observation of the Kolmogorov
law is only obtained in presence of the normal component--see \cite%
{Maurer1998}. There is as yet no direct experimental evidence relating to
the spectrum of the turbulent energy at very low temperatures (see \cite%
{Vinen2002}). Measuring the fluctuation of the VLD \ (see, \cite{Roche2007},
\cite{Bradley2008}) gave results, which are, probably, inconsistent with the
quasi-classical behavior of QT. As for numerical results, there are several works, which
demonstrate the dependence $E(k)\propto \,k^{-5/3}$. There are
both the works, based on the vortex filament methods (VFM) \cite{Araki2002,Kivotides2002,Kivotides2001c}
and works using GPE \cite{Nore1997,Nore1997a}, \cite{Kobayashi2005},\cite%
{Sasa2011}.

The most common view of quasi-classical turbulence is the model of vortex
bundles. The point is that the quantized vortices have a fixed core radius,
so they don't possess the very important property of classical turbulence
-- stretching vortex tubes with a decrease in the core size. The latter is
responsible for the turbulence energy cascade from large scales to the small
scales. Collections of near-parallel quantized vortices (vortex bundles) do
possess this property, so the idea that the quasi-classical turbulence in
quantum fluids is realized via vortex bundles of different sizes and
intensities (number of threads ) seems quite natural. However the concept of
the bundle structure does not explain appearance of Kolmogorov type spectrum
$E(k)\propto \,k^{-5/3}$, since the usual uniform vortex array just
generates the coarse grained solid body rotation.

In the work we study nonuniform vortex arrays, whose structure is determined
by the collapsing vortex dynamics.

\section {Uniform vortex array}

The energy of the vortex tangle, expressed via vortex filaments elements $\mathbf{%
s}(\xi _{j})$ ($\xi _{j}$ is the label variable of  $j$- loop ) is defined
as (see \cite{Nemirovskii1998},\cite{Nemirovskii2002},\cite{Nemirovskii2013a})

\begin{equation}
E(k)=\frac{\rho _{s}\kappa ^{2}}{(2\pi )^{2}}\sum_{i,j}\int%
\limits_{0}^{L_{i}}\int\limits_{0}^{L_{j}}\mathbf{s}_{j}^{\prime }(\xi
_{i})\cdot \mathbf{s}_{j}^{\prime }(\xi _{j})d{\xi }_{i}d{\xi }_{j}\frac{%
\sin (k\left\vert \mathbf{s}(\xi _{j})-\mathbf{s}(\xi _{j})\right\vert )}{%
k\left\vert \mathbf{s}(\xi _{i})-\mathbf{s}(\xi _{j})\right\vert }
\label{E(k) spherical}
\end{equation}

For anisotropic situations, formula (\ref{E(k) spherical}) is understood as
an angular average, but one has  to treat this formula with precaution (see
\cite{Nemirovskii2015}). Thus, for calculation of the energy spectrum $E(k)$%
\ of the 3D velocity field, induced by the vortex filament we
need to know the exact configuration $\{s(\xi )\}$\ of vortex lines.

Let's study the question, what is the energy spectrum of 3D flow induced by the array of
vortex filaments, imitating the bundle. First we consider a set of straight
vortex filaments forming the square lattice $\bigcup s_{i}(\xi )=\bigcup
(x_{i},y_{j},z)$. Points $x_{i},y_{j}$\ are coordinates for vortices on the $%
xy$-plane, indices $i,j$\ runs from $1$\ to $N$\ . The neighboring lines are
separated by distance $b$, i.e., $x_{i+1}-x_{i}=b$. In case of different
straight lines we have to perform integration between different lines
and $\left\vert \mathbf{s}_{1}(\xi _{1})-\mathbf{s}_{2}(\xi _{2})\right\vert
=\sqrt{(x_{1i}-x_{2i})^{2}+(y_{1j}-y_{2j})^{2}+(z_{i}-z_{j})^{2}}=\sqrt{%
d_{ij}^{2}+(z_{i}-z_{j})^{2}}$  where $d_{ij}=\sqrt{%
(x_{1i}-x_{2i})^{2}+(y_{1j}-y_{2j})^{2}}$ distances between vortices on the $%
xy$ -plane. Then equation (\ref{E(k) spherical}) can be rewritten as
\begin{eqnarray}
\frac{E(k)}{{\rho }_{s}{\kappa }^{2}L}=\frac{1}{4\pi k}%
\sum_{i_{1},i_{2}=1}^{N}\sum_{j_{1},j_{2}=1}^{N}\int\limits_{0}^{L}\int%
\limits_{0}^{L}\frac{\sin (k\sqrt{d_{ij}^{2}+(z_{i}-z_{j})^{2}})}{(\sqrt{%
d_{ij}^{2}+(z_{i}-z_{j})^{2}})}\;d(z_{1}-z_{2})
\end{eqnarray}%
Integral over $z$ is in the table by Ryzhik \& Gradshtein (3.876) (see \cite%
{Gradshteyn1980})
\begin{equation}
\frac{E(k)}{{\rho }_{s}{\kappa }^{2}L}=\frac{1}{4\pi k}%
\sum_{i_{1},i_{2}=1}^{N}\sum_{j_{1},j_{2}=1}^{N}J_{0}(kd_{ij}).
\label{Risto lattice}
\end{equation}%
Thus, determination of the spectrum on the basis (\ref{Risto lattice})
should be done with the use of the quadruple summation (over $%
(x_{i},x_{j},y_{i},y_{j})$), which requires large computing resources.
Clear, however, that for very small $k$, which corresponds to very large
distance, the whole array can be considered as large single vortex with the
circulation $N^{2}\kappa $. Accordingly, the spectrum (per unit height)
should be ($\rho _{s}N^{4}\kappa ^{2}/4\pi )k^{-1}$. For large $k$, which
corresponds to very small distance from each line, the spectrum (per unit
height) should be $(\rho _{s}\kappa ^{2}/4\pi )k^{-1}$ as for the single
straight vortex filament. In the intermediate
region $kb<<1$, and $Nkb>>1$\ (this condition implies that inverse wave
number $k^{-1}$\ is larger than the intervortex space between neighboring lines, but
smaller then the size of the whole array $Nb$), we can replace the quadruple
summation by the quadruple integration with infinite limits. This procedure
corresponds that we exclude the fine-scale motion near each of vortex, and
are interested in the only large-scale, coarse-grained motion. After obvious
change of variables $x_{i}\rightarrow kx_{i},\ y_{i}\rightarrow ky_{i}$\
etc. we get that the whole integral should scale as $1/k^{4}$, and
accordingly $E(k)\propto 1/k^{5}$ (compare with \cite{Nowak2012}). As it is shown in \cite{Nemirovskii2013a}, the
velocity $v(\mathbf{r})$\ scales as $\mathbf{r}^{1}$.
Thus, the uniform vortex array creates the course-grained motion, which is
rotation (velocity is proportional to the distance from center), as it
should be. Moreover, the coefficient is proportional to $\kappa /2b^{2}$,
which corresponds to the Feynman rule. Concluding this subsection we state that the uniform
vortex bundles do not generate the Kolmogorov spectra.

\section {Vortex lines breaking}

\begin{figure}
\includegraphics[width=6cm]{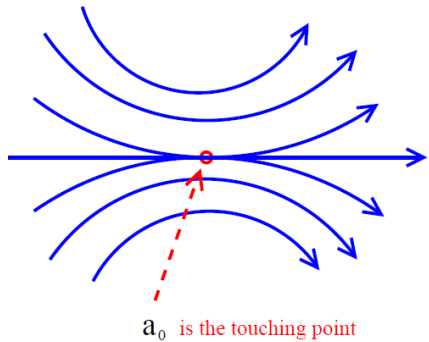}
\caption{(Color online)Schematic picture illustrating the vortex bundle collapse \cite{Kuznetsov2013}. The regular initially distribution
of vorticity spontaneously concentrates, collapsing in some point $\mathbf{a}_{0}$
and forming the singular structure.  }
\label{breaking}
\end{figure}

Currently, in classic hydrodynamics, the highly important topic - the role
of hydrodynamic collapses in the formation of turbulent spectra -
is being intensively discussed (See e.g., \cite{Kuznetsov2000}, \cite%
{Kerr2013}, \cite{Nemirovskii1982}). Briefly, this phenomenon can be described as spontaneous
infinite growth of the vorticity field $\mathbf{\omega }=\nabla \times
\mathbf{v}$ with formation of singularity in $\omega (r)$. In
particular, in the continuously distributed vortex field the vortex lines
(not quantized vortex filaments, just hydrodynamic vortex lines!) start to
accumulate at some points $ a_{0}$ forming singular distribution \ $%
\omega (r)\sim r^{-2/3}$ as it is illustrated in Fig. \ref{breaking}. The latter results in the increment for velocity
field \ $\left\langle \mathbf{v}(\mathbf{r+\delta r)}\right\rangle \sim
r^{1//3}$, which, in turn results it to the Kolmogorov spectrum $%
E(k)\propto \,k^{-5/3}$.
In classical hydrodynamics this scenario is known as the vortex breaking. This phenomenon  was analyzed in series of papers by Kuznetsov with coauthors (
see/ e.g. \cite{Kuznetsov2000},\cite{Agafontsev2015} and references
therein)in the framework of the integrable incompressible hydrodynamic model
with the Hamiltonian $\int |\mathbf{\omega }|d\mathbf{r}$.
Studying Euler equations in terms of vorticity (also known as Helmholtz's
vorticity equations)
\begin{equation}
\frac{\partial \mathbf{\omega }}{\partial t}=\nabla \times \,(\mathbf{v}%
\times \mathbf{\omega }),  \label{dw/dt classical}
\end{equation}%
Kuznetsov concluded that in the vicinity of the touching point $\mathbf{a}%
_{0}$ the maximum value of vorticity $\omega _{\max }$\ develops in the
blow-up manner%
\begin{equation}
\omega _{\max }\propto \frac{1}{t^{\ast }-t},  \label{omagamax}
\end{equation}%
with the approaching the infinity at some time $t^{\ast }$. The domain of
vorticity is not isotropic, it has a pancake structure. The main dependence
of vorticity field is connected with the transverse to the bundle direction $%
\rho _{\perp },$ and  $%
\omega (\rho _{\perp })$ scales as
\begin{equation}
\omega \propto \frac{1}{\rho _{\perp }^{2/3}}.  \label{omegafield}
\end{equation}%

The similar consideration can be applied for quantum vortices. It, however,
can be done only in particular case of vortex bundles, when the quantum vortex
filaments form a
near-parallel structure. In this case The coarse-grained hydrodynamic
equations for the superfluid vorticity are obtained from the Euler equation
for the superfluid velocity $\mathbf{v}_{s}\mathbf{=}\left\langle \mathbf{v}%
_{s}\right\rangle $ after averaging over the vortex lines . The
coarse-grained hydrodynamic of the vortex bundles is studied by many authors
(see e.g., \cite{Sonin1987},\cite{HENDERSON2000},\cite{Holm2001},\cite%
{Jou2011}), but the basis of these studies was the hydrodynamics of rotating
superfluids, or the Hall-Vinen-Bekarevich-Khalatnikov (HVBK) model (see
e.g., book \cite{Khalatnikov1965}). In the vortex bundles, the coarse-grained vorticity
field of $\omega _{s}$,  and the 2D vortex
line density $\mathcal{L}$ (which coincides with the areal density in plane
perpendicular to the bundle) are related to each other by means of the
Feynman's rule, $\omega _{s}=\kappa \mathcal{L}.  \label{bundle omega}$. In terms
HVBK the dynamics of this vorticity obeys the following equation (see
\cite{Khalatnikov1965})
\begin{equation}
\frac{\partial \mathbf{\omega }_{s}}{\partial t}=\nabla \times \lbrack
\mathbf{v}_{L}\times \mathbf{\omega }_{s}],  \label{dw/dt HVBK}
\end{equation}%
where $\mathbf{v}_{L}$ is the velocity of lines,

\begin{equation}
\mathbf{v}_{L}\mathbf{=v}_{s}\mathbf{+}\alpha \left[ \mathbf{\hat{\omega}}%
_{s}{\times }\left( \mathbf{v}_{n}-\mathbf{v}_{s}\right) \right] .
\label{V_L}
\end{equation}

It is easy to see that in case of zero temperature, when mutual friction
vanishes $\alpha =0$, and taking into account that vortex lines move with the averaged
velocity $\mathbf{v}_{s}\mathbf{=}\left\langle \mathbf{v}_{s}\right\rangle ,$
the dynamics of macroscopic (or the coarse-grained) vorticity is identical to
the dynamics of classical field, therefore all, stated above  conclusions concerning the collapse of vorticity are valid for quantum fluids.

\section {Noninform lattice}

Let's now consider the nonuniform vortex bundle. To model this situation we
just can choose that the distance $b$ between lattice points (see Sec. 2) is not constant, but depends on the numbers $i,j$ of the cell
nodes. We have to realize that the problem of the spontaneous formation of
vortex bundles is only on the stage of discussion so far, and there is no
ideas concerning an exact arrangement of these bundles. We will choose the power
law dependence for the distance between the lattice points.
\begin{equation}
x_{i+1}-x_{i}=b_{0}i^{\lambda },\ y_{j+1}-y_{j}=b_{0}j^{\lambda }.
\label{nonuniform}
\end{equation}%
We do not ascertain the quantity $\lambda $, it is free parameter of our
approach. Under condition (\ref{nonuniform}) the expression (\ref{Risto
lattice}) turns into
\begin{equation}
\frac{E(k)}{{\rho }_{s}{\kappa }^{2}L}=\frac{1}{4\pi k}\sum_{i=1}^{N}%
\sum_{j=1}^{N}J_{0}(kb_{0}i^{\lambda })
\end{equation}

That means that while changing the summation by integration we have to put $%
x_{i}\rightarrow k^{1/\lambda }x_{i},\ y_{i}\rightarrow k^{1/\lambda }y_{i}\
$(instead of the change of variables $x_{i}\rightarrow kx_{i},\
y_{i}\rightarrow ky_{i}$\, made in Sec. 2). As a result we get, that the
whole integral should be scaled as $1/k^{1+4/\lambda }$. It is easy to see that
when $\lambda $ $=6$, the spectrum $E(k)\propto k^{-5/3}$.

Let's find the $2D$ density of vortices on the $xy$ plane under condition (\ref%
{nonuniform}), or, according to the Feynman rule, the distribution of
vorticity $\omega (r)$. In the "space" of indices $\{i,j\}$ vortices are
distributed uniformly (one vortex per lattice site $\{i,j\}$), but since the
distances between the sites vary, the distribution of \ vortices in the real
$xy$ space is nonuniform. Let us consider "the ring" of radius from $I$ to $%
I+\Delta I$ in $\{i,j\}$. Then, the number of points $\Delta N$ in ring is
just $2\pi I\Delta I$, the radius of ring in real $xy$ space is $%
r=b_{0}I^{\lambda }$, and the thickness of ring is $\Delta r=b_{0}\lambda
I^{\lambda -1}\Delta I$. From these relations it follows that the real $n(r)$ scales
with $r$ as
\begin{eqnarray}
n(r)=\frac{\Delta N}{\Delta r}\propto \frac{1}{r^{1-2/\lambda }}.
\end{eqnarray}

If $\lambda $ $=6$, then $\kappa n(r)=\omega (r) \propto r^{-2/3}$, as it should be for the
classical turbulence \cite{Kuznetsov2000},\cite{Agafontsev2015}.

\section{Conclusions}

Summarizing, it can be concluded that the 3D energy spectrum $E(k)$
close to the Kolmogorov dependence $\propto k^{-5/3}$\,
which was observed in many numerical simulations on the superfluid
turbulence \cite{Araki2002}-\cite{Sasa2011}, can appear from the collapsing
vortex bundle.

The work was supported by grant 15-02-05366  from RFBR (Russian Foundation of
Fundamental Research)



%
%

\end{document}